\documentclass[lettersize,journal]{IEEEtran}
\usepackage{amsmath,amsfonts,amssymb}
\usepackage{algorithmic}
\usepackage{algorithm}
\usepackage{array}
\usepackage[caption=false,font=normalsize,labelfont=sf,textfont=sf]{subfig}
\usepackage{textcomp}
\usepackage{stfloats}
\usepackage{url}
\usepackage{verbatim}
\usepackage{graphicx}
\usepackage{cite}
\usepackage{xcolor}
\usepackage{booktabs}
\usepackage{multirow}
\usepackage{svg}
\usepackage{pgfplots}
\usepackage{tikz}
\pgfplotsset{compat=newest}

\definecolor{globecomblue}{RGB}{0, 83, 155}
\definecolor{globecomred}{RGB}{204, 51, 0}
\definecolor{globecomcyan}{RGB}{0, 150, 200}

\pgfplotsset{
	mybarstyle/.style={
		ybar, 
		width=\columnwidth, 
		height=5.5cm, 
		ymin=50, ymax=102, 
		ylabel={Accuracy (\%)}, 
		ylabel style={font=\footnotesize, yshift=-0.5em},
		xtick=data, 
		xticklabel style={rotate=45, anchor=east, font=\scriptsize, align=right},
		nodes near coords,
		nodes near coords align={vertical},
		every node near coord/.append style={font=\tiny, /pgf/number format/fixed, /pgf/number format/precision=1, yshift=0.2em},
		legend style={at={(0.5,1.15)}, anchor=south, legend columns=-1, font=\footnotesize, draw=none, fill=none},
		ymajorgrids=true, 
		grid style=dashed, 
		bar width=12pt, 
		enlarge x limits=0.2, 
		tick label style={font=\footnotesize} 
	},
	groupedbar/.style={
		mybarstyle,
		bar width=9pt, 
	}
}

\begin{document}
	
	\title{Robust Cross-Domain WiFi Fall Detection via Physics-Driven Attention-Enhanced Transformers}
	
	\author{Yingzhe Wang,
		Cunhua Pan,~\IEEEmembership{Senior Member,~IEEE, } 
	    Ruijing Liu,~\IEEEmembership 
	    Shaokai Li,
		Hong Ren,~\IEEEmembership{Member,~IEEE, } 
		
		Kezhi Wang, ~\IEEEmembership{Senior Member,~IEEE, }
		Jiangzhou Wang,~\IEEEmembership{Fellow,~IEEE}
		\thanks{Yingzhe Wang, Cunhua Pan, Ruijing Liu, Shaokai Li, and Hong Ren are with the National Mobile Communications Research Laboratory, Southeast University, Nanjing 210096, China (e-mail: 220250908@seu.edu.cn; cpan@seu.edu.cn; 230258158@seu.edu.cn; shaokaili@seu.edu.cn; hren@seu.edu.cn).}%
		\thanks{Kezhi Wang is with the Department of Computer Science, Brunel University
			London, UB8 3PH Uxbridge, U.K. (e-mail: kezhi.wang@brunel.ac.uk).}
		\thanks{Jiangzhou Wang is with the National Mobile Communications Research Laboratory, Southeast University, Nanjing 210096, China, and also with the Pervasive Communication Research Center, Purple Mountain Laboratories, Nanjing 211111, China (e-mail: j.wang@seu.edu.cn).}%
		}
	
	
	
	\maketitle
	
	\begin{abstract}
		Device-free fall detection utilizing WiFi Channel State Information (CSI) has emerged as a promising, privacy-preserving solution for elderly health monitoring in the Internet of Things (IoT) era. However, existing deep learning approaches suffer from severe performance degradation when deployed in unseen environments due to static background overfitting and Non-Line-of-Sight (NLoS) signal attenuation. To address these critical bottlenecks, we propose a robust, domain-generalizable framework featuring a novel Attention-Enhanced CNN-Transformer hybrid architecture. First, we design a physics-driven \textbf{Dynamic Variance Gate (DVG)} to dynamically calculate local temporal variance, acting as a soft-attention mask that eliminates static environmental DC components while amplifying dynamic human motion. Second, we introduce a Physics-Aware Data Augmentation strategy to force the network to learn invariant morphological signatures rather than environment-specific noise. Furthermore, a Convolutional Block Attention Module (CBAM) is integrated to refine spatiotemporal features prior to Transformer-based sequence modeling. Extensive cross-domain evaluations across four distinct indoor environments demonstrate that our method achieves 97.6\% accuracy in NLoS scenarios and 98.8\% in completely unseen environments without target-domain fine-tuning. Finally, we deploy the proposed framework on an edge computing system equipped with commercial WiFi NICs. Real-world live inference field tests confirm the system's robustness against unseen environmental layouts and its capability for continuous, low-latency whole-home safety monitoring.
	\end{abstract}
	
	\begin{IEEEkeywords}
		Fall detection, WiFi sensing,   Channel State Information (CSI), Deep Learning,Transformer, Domain Generalization
	\end{IEEEkeywords}
	
	\section{Introduction}
	\IEEEPARstart{F}{alls} are among the most frequent and dangerous accidents affecting older adults, especially those living alone in increasingly aging societies. Beyond the immediate injury, a fall often triggers a prolonged period of immobility, commonly referred to as the ``long lie,'' which can substantially increase the risks of dehydration, rhabdomyolysis, hospitalization, and mortality if timely assistance is unavailable \cite{b1}. From the perspective of smart healthcare and the Internet of Things (IoT), this makes continuous and reliable fall detection a foundational capability for aging-in-place services. In practice, however, a useful fall detection system must satisfy several stringent requirements simultaneously: it should operate continuously, preserve privacy, work under poor lighting and occlusion, require minimal user participation, and remain robust across different rooms, furniture layouts, and deployment conditions.
	
	Conventional sensing modalities struggle to satisfy these requirements at the same time. Wearable devices can capture acceleration and posture information effectively, but their performance depends heavily on user compliance. Elderly users may forget to wear or charge them, reject them due to discomfort, or be unable to trigger an alarm after losing consciousness. Vision-based systems offer strong perceptual capability, yet they raise severe privacy concerns in sensitive indoor areas such as bedrooms and bathrooms and remain vulnerable to illumination changes, occlusions, and blind spots. Radar-based approaches alleviate some privacy concerns, but dedicated hardware raises deployment cost and their coverage is often constrained by line-of-sight and installation density. These limitations motivate a strong need for a privacy-preserving, low-cost, and device-free sensing solution that can be seamlessly integrated into real homes.
	
	WiFi sensing based on Channel State Information (CSI) has emerged as a compelling candidate for this purpose because it can reuse ubiquitous communication infrastructure to perceive human motion in a passive and unobtrusive manner \cite{b2,b34}. CSI provides fine-grained channel responses across multiple subcarriers, and its amplitude and phase variations implicitly encode how human motion perturbs the multipath propagation structure. This property has enabled a broad class of device-free sensing applications, including human activity recognition, gesture recognition, localization, respiration monitoring, and fall detection.
	
	In the specific context of fall detection, a growing body of work has confirmed that commodity WiFi signals carry sufficient motion information to distinguish falls from daily activities without cameras or body-worn sensors \cite{b16,b15,b18,b11,b17,b3,b29}, and the adoption of attention mechanisms and Transformer-style architectures has further improved spatiotemporal feature quality \cite{b6,b7,b5,b8,b9}. Despite these advances, many reported results have been obtained in controlled environments where training and test conditions share the same room, transceiver configuration, and background structure. This evaluation protocol conceals a fundamental mismatch: a fall detection system for elderly care must operate reliably across an effectively unbounded variety of home layouts, furniture arrangements, and signal propagation conditions that no finite training set can fully represent. Bridging this gap without per-home data collection or re-training is not merely a performance objective but an absolute prerequisite for a practically deployable IoT safety system, and it remains unresolved by the majority of existing approaches.
	
	The first challenge is domain-dependent background bias. WiFi sensing is physically inseparable from the environment in which it operates: the received CSI is jointly determined by room geometry, wall materials, furniture placement, and transceiver positions, all of which collectively shape the static multipath structure of the channel. A deep learning model trained in one room inevitably absorbs this multipath fingerprint as a discriminative feature, learning to exploit environment-specific amplitude and phase patterns that correlate with falls in the training setting but do not transfer. The result is a systematic failure mode rather than a gradual accuracy reduction: such a model may achieve near-perfect accuracy within the training room while producing clinically unacceptable false-negative rates in a structurally different bedroom, kitchen, or corridor. Resolving this through target-domain adaptation is impractical in elderly care, where supervised calibration data cannot be safely or logistically collected in every new home, and where the deployment value of the system depends entirely on zero-configuration operation. Although prior cross-domain methods \cite{b19,b20,b21,b22,b13,b10} have made meaningful progress, many still require target-domain samples or multi-stage training pipelines that conflict with plug-and-play IoT deployment.
	
	The second challenge arises from Non-Line-of-Sight (NLoS) and through-wall propagation \cite{b25,b31}. In realistic homes, falls may occur behind furniture, around corners, or across walls rather than within a clear propagation path. Under such conditions, the high-frequency Doppler components associated with sudden body impact are partially absorbed and scattered before reaching the receiver, producing CSI signatures that are weaker and spectrally smoother than their Line-of-Sight (LoS) counterparts \cite{b26,b35}. This attenuation can be understood as a form of environment-induced low-pass filtering applied to the motion-induced channel variation. As a result, a model calibrated on clear LoS recordings may systematically fail to detect falls that occur in obstructed areas---precisely those areas, such as bathrooms and narrow corridors, where falls are most frequent and most dangerous \cite{b27}. The NLoS problem thus represents not merely a performance degradation but a coverage gap that undermines the safety guarantee of any deployed system.
	
	The third challenge concerns spatiotemporal modeling. Unlike a static posture classification task, a fall is an ordered multi-stage event: a pre-fall postural instability, a rapid descent lasting roughly 0.3--0.5 seconds, a high-energy floor impact, and a prolonged period of post-fall stillness. Correct detection depends not only on recognizing the CSI signature of each individual phase but on capturing the causal temporal relationship among them. A model that perceives only local time-frequency patterns risks misclassifying a rapid sit-down, an energetic forward lean, or a sudden crouch as a fall, and may confuse a post-fall subject lying motionless with someone who has simply sat down to rest---both error types carry immediate safety consequences in a single-occupant elderly monitoring context. Yet simply expanding the model's temporal receptive field is insufficient if the input CSI is contaminated by static environmental clutter: without a physics-grounded front-end that isolates dynamic motion components from background interference, a sequence model risks learning the temporal autocorrelation of environmental noise rather than the true morphology of a fall. For always-on edge deployment, these requirements must furthermore be satisfied within a strict latency budget that precludes reliance on large cloud-based models.
	
	To address these three challenges, we propose a robust cross-domain WiFi fall detection framework that explicitly embeds wireless propagation priors into both signal preprocessing and model learning, enabling reliable zero-shot generalization without target-domain adaptation. As illustrated in Fig.~\ref{smm}, the proposed system integrates a physics-driven preprocessing pipeline with an attention-enhanced CNN-Transformer architecture. The main contributions of this paper are summarized as follows:
	
	\begin{figure}[t]
		\centering
		\includegraphics[width=0.9\columnwidth]{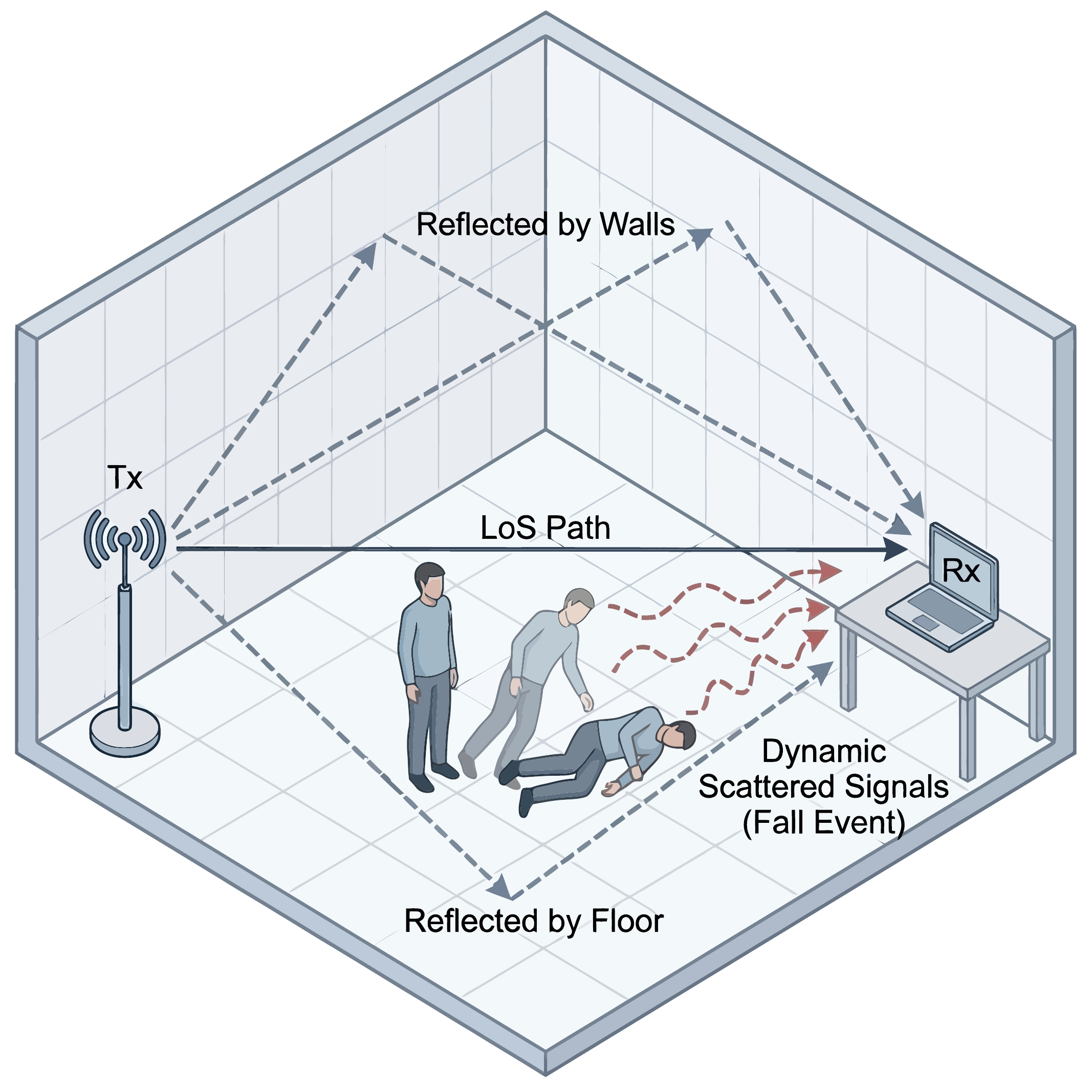}
		\caption{Overview of the proposed physics-aware, attention-enhanced WiFi fall detection system.}
		\label{smm}
	\end{figure}

	\begin{enumerate}
		\item \textbf{Physics-driven environment decoupling via Dynamic Variance Gate (DVG):} We propose a novel DVG module that computes local temporal variance over CSI streams and transforms it into a soft attention mechanism. By suppressing static environmental clutter and enhancing motion-sensitive regions at the input stage, DVG decouples fall signatures from room-specific background bias before deep feature extraction.
		
		\item \textbf{Attention-enhanced CNN-Transformer hybrid architecture:} We integrate a Convolutional Block Attention Module (CBAM) between an EfficientNet-B0 backbone \cite{b24} and a Transformer encoder. This dual-attention design adaptively highlights critical spatiotemporal regions and models the long-range temporal dependencies inherent in multi-stage fall events within a single unified architecture.
		
		\item \textbf{Physics-Aware Data Augmentation for cross-domain and NLoS robustness:} We introduce an augmentation strategy grounded in wireless propagation physics, including Rician-style noise perturbation, distance-aware amplitude scaling, temporal shifting, and low-pass spectral degradation to simulate NLoS attenuation. This forces the network to learn invariant fall morphology rather than environment-specific artifacts, without requiring any target-domain data.
		
	    \item  The proposed framework achieves 97.6\% accuracy in challenging NLoS scenarios and maintains a resilient 98.8\% accuracy in completely unseen environments without requiring target-domain fine-tuning. Crucially, real-world live inference field tests on an edge computing system confirms its low-latency performance and practical robustness against physical environmental changes.
	\end{enumerate}
	
	The remainder of this paper is organized as follows. Section~II reviews related work on WiFi-based activity sensing and cross-domain generalization. Section~III details the proposed physics-aware preprocessing pipeline and the attention-enhanced hybrid network architecture. Section~IV presents the experimental setup and cross-domain evaluation results. Section~V concludes the paper and discusses future directions for practical IoT-enabled elderly care.
	
	\section{Related Work}
	\subsection{WiFi-Based Activity and Fall Detection}
	In recent years, Device-Free Passive (DFP) sensing using CSI has evolved from handcrafted statistical features to deep spatiotemporal learning. Early representative fall detection systems such as RT-Fall \cite{b16} and FallDeFi \cite{b15} demonstrated that commodity WiFi signals can support contactless fall monitoring in indoor environments. Subsequent studies improved feature representation and environment robustness through spectrogram-image analysis \cite{b18}, environment-independent motion feature learning \cite{b11}, spatial angle-of-arrival modeling \cite{b17}, and explainable robust classification \cite{b3}. These works established WiFi fall detection as a viable privacy-preserving alternative to camera-based or wearable solutions, but they largely rely on handcrafted features or rule-based pipelines that limit representational capacity, and most are evaluated within a single controlled environment that does not reflect the deployment diversity of real homes.

	To improve feature expressiveness and temporal modeling capability, subsequent research turned to deep learning architectures. CNNs became a common choice for extracting local spatiotemporal patterns from CSI-derived spectrograms \cite{b4,b18}, but local convolution alone is insufficient for modeling the complete multi-stage temporal evolution of a fall. Hybrid models addressed this by incorporating recurrent units and attention mechanisms: CNN-GRU-AttNet \cite{b6} combines convolutional layers with gated recurrent units and self-attention to enhance sequence understanding, while Fall-attention \cite{b7} employs RNN-based encoding with attention-guided aggregation to handle fall-adjacent confounding activities. More recently, pure Transformer-based WiFi sensing models have demonstrated superior capability for capturing long-range dependencies and global semantic interactions across entire event windows \cite{b5,b8,b9,b33}. However, these architectures are generally evaluated in single-domain settings and do not systematically address robustness to domain shift or NLoS degradation. Motivated by these observations, our framework integrates local convolution, dual attention refinement, and Transformer-based temporal reasoning within a unified architecture explicitly designed for cross-domain fall detection.
	
	\subsection{Cross-Domain Generalization in WiFi Sensing}
	Despite achieving high accuracy in controlled settings, WiFi-based sensing models often suffer from severe performance degradation under domain shifts caused by varying room layouts, transceiver placements, user characteristics, and physical obstacles. This problem has been studied across several WiFi sensing tasks. CrossSense \cite{b19} improves cross-site sensing by generating synthetic environment-specific samples through a roaming model, while TL-Fall \cite{b20} reuses source-domain knowledge via transfer learning to reduce the labeled data required in a new environment. Other studies have explored semi-supervised cross-location adaptation \cite{b21}, iterative soft labeling combined with domain alignment \cite{b22}, adversarial generative feature mapping \cite{b13}, and multi-source domain generalization \cite{b28} that aggregates diverse training environments to improve zero-shot transferability \cite{b10}. More recently, Wi-CBR \cite{b12} proposed a salient-aware adaptive sensing strategy for cross-domain behavior recognition, further highlighting the community's growing interest in deployment-robust WiFi systems.
	
	Although these methods significantly advance cross-domain WiFi sensing, they carry deployment assumptions that are difficult to satisfy in the fall detection scenario specifically. While few-shot learning techniques can alleviate retraining overheads \cite{b32}, methods that depend heavily on target-domain calibration data still require the deployment site to be instrumented and sampled in advance---an impractical requirement when the goal is zero-configuration installation in a private elderly home. Methods that rely on multiple training stages or sophisticated adversarial objectives increase system complexity and may introduce training instability in the low-data regimes typical of fall datasets. Moreover, the majority of these works are validated on activity recognition or gesture datasets in which inter-class differences are large and well-separated; fall detection presents a more challenging generalization problem because the distinction between a genuine fall and a confounding motion such as a rapid sit-down or a stumble-recovery is subtle and critically dependent on the full temporal evolution of the event. For plug-and-play elderly monitoring, a more robust path is to prevent background bias from being learned in the first place, rather than correcting for it after the fact. To this end, our framework combines instance-wise normalization, a physics-driven Dynamic Variance Gate, and wireless-aware data augmentation to reduce background dependency and improve robustness to both unseen room layouts and NLoS attenuation, without requiring any target-domain data.
	
	\section{SYSTEM DESIGN}
\begin{figure*}[htbp]
		\centering
		\includegraphics[width=\textwidth]{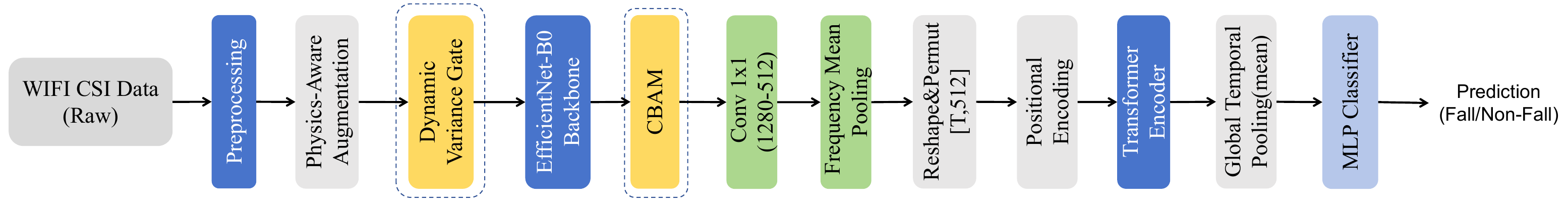}
		\caption{The full-network architecture for FallDetection.}
		\label{fig:architecture}
		\vspace{1.5em}
		\includegraphics[width=\textwidth]{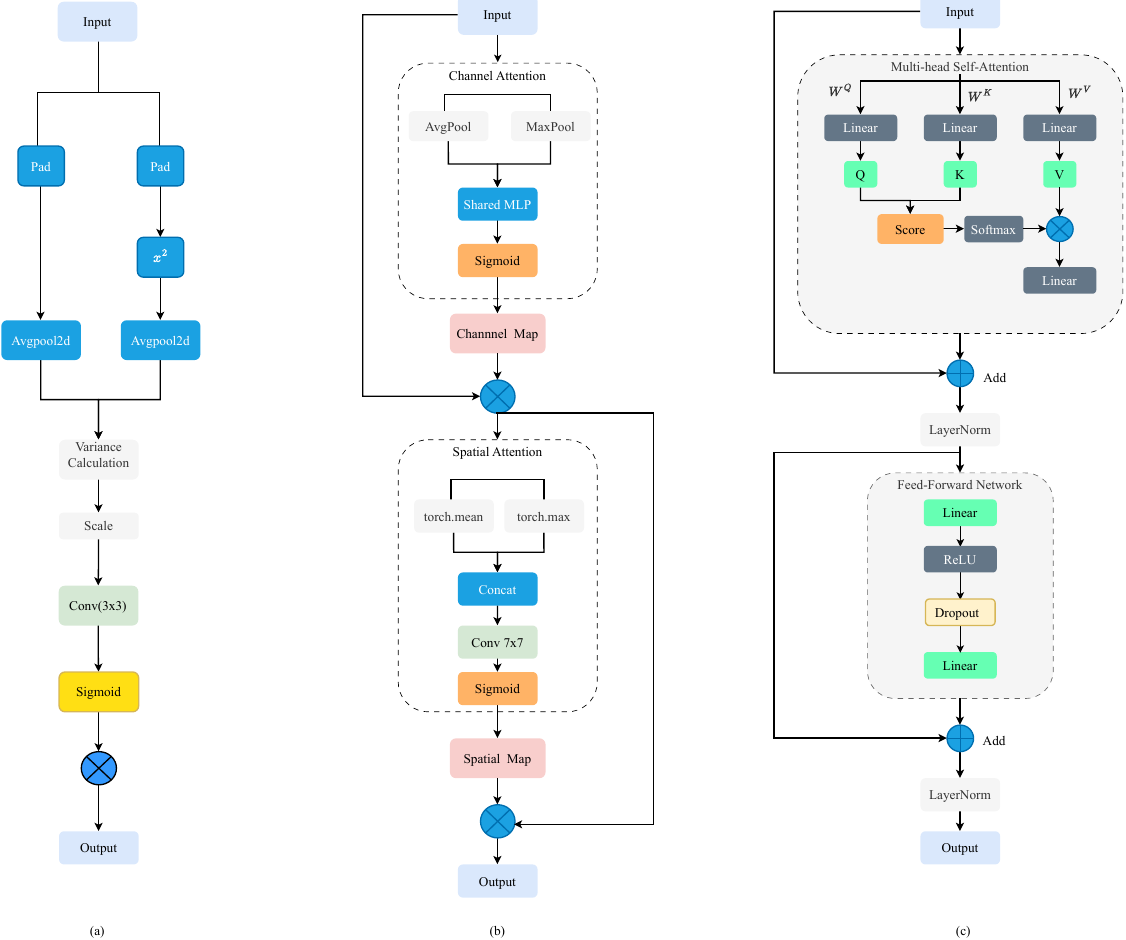}
		\caption{(a)Detailed view of the Dynamic Variance Gate module. (b)Detailed view of the CBAM module. (c)Detailed view of Transformer Encoder module.}
		\label{fig:divide}
	\end{figure*}

	As illustrated in Fig. \ref{fig:architecture} and Fig. \ref{fig:divide}, the proposed framework is designed as a physics-aware and deployment-oriented pipeline for cross-domain WiFi fall detection. The overall system contains four tightly coupled components: 
	1) a lightweight preprocessing front-end that converts raw CSI streams into fixed-size instance-normalized tensors; 
	2) a physics-aware augmentation strategy applied during training to improve robustness to deployment shifts such as distance variation and NLoS attenuation; 
	3) a Dynamic Variance Gate (DVG) preceded by channel standardization, which suppresses static environmental clutter and highlights motion-sensitive regions; 
	and 4) a hybrid CNN-Transformer backbone with CBAM-based feature refinement for spatiotemporal representation learning.
	
	\subsection{Physics-Aware Preprocessing}
	The input to our model is a continuous CSI stream collected by commodity WiFi transceivers. Before deep feature extraction, the stream must be converted into fixed-size samples that preserve the temporal structure of a fall while remaining efficient enough for real-time deployment. To this end, we adopt a lightweight preprocessing pipeline consisting of sliding-window segmentation, temporal down-sampling, tensor reorganization, and instance normalization. Unlike preprocessing-heavy schemes that rely on handcrafted denoising or frequency-domain image generation, our front-end is deliberately kept simple so that environment suppression can be handled adaptively by the learnable DVG in the next stage.
	
	\subsubsection{Sliding-Window Segmentation}
	A fall is a multi-stage event rather than an instantaneous impulse. It typically includes posture transition, rapid descent, impact, and a short post-fall period. Therefore, the raw CSI stream is partitioned using a fixed-length sliding window so that each sample captures a complete temporal context. Specifically, CSI packets are collected at 1000 packets/s, and a 5-s window is used to extract one local segment:
	\begin{equation}
		X_{raw} \in \mathbb{R}^{T_0 \times S_0},
	\end{equation}
	where $T_0 = 5000$ and $S_0$ denotes the number of CSI streams before channel reorganization. The overlap between neighboring windows can be adjusted according to the latency and throughput budget of the deployment platform.
	
	\subsubsection{Temporal Down-Sampling}
	Although the CSI is captured at 1 kHz, most discriminative motion energy of fall events is concentrated in relatively low temporal frequencies. Processing the full 5000-packet sequence therefore introduces redundant samples and unnecessary computational cost. We down-sample each segmented sample by a factor of 8:
	\begin{equation}
		X_{ds} = \text{DownSample}(X_{raw}, r=8),
	\end{equation}
	where $X_{ds} \in \mathbb{R}^{625 \times S_0}$. The resulting effective sampling rate is 125 Hz, which is sufficient to preserve the dominant fall-related dynamics while substantially reducing memory usage and inference latency.
	
	\subsubsection{Tensor Reorganization}
	After down-sampling, the CSI segment is reorganized into a tensor structure compatible with the image-style backbone network. Under the Intel 5300 CSI setting used in this work, one transmitting antenna and three receiving branches yield 90 CSI streams in total. The segment is therefore reshaped from a matrix of size $625 \times 90$ into
	\begin{equation}
		X_{re} \in \mathbb{R}^{3 \times 625 \times 30},
	\end{equation}
	where the first dimension indexes the three receive branches and each branch contains the temporal responses of 30 subcarriers. This representation preserves antenna diversity while exposing local time-subcarrier patterns to the subsequent convolutional encoder.
	
	\subsubsection{Instance and Channel Normalization}
	In cross-domain deployments, the absolute CSI amplitude is strongly influenced by room geometry, transceiver separation, furniture layout, and blockage conditions.
	If global training-set statistics are used for normalization, these deployment-specific biases may be retained and inadvertently learned by the classifier.
	To reduce this effect, we first normalize each sample independently. For a reorganized tensor $X_{re} \in \mathbb{R}^{C \times T \times S}$, the instance-normalized sample is computed as:
	\begin{equation}
		X'_{c,t,s} = \frac{X_{re}(c,t,s) - \min(X_{re})}{\max(X_{re}) - \min(X_{re}) + \epsilon},
	\end{equation}
	where $\epsilon$ is a small constant for numerical stability. 
	\subsection{Training-Time Physics-Aware Augmentation}
	Cross-domain robustness cannot rely only on architectural design; it must also be reinforced during training. Therefore, in addition to the inference-time preprocessing and gating pipeline above, we apply a physics-aware augmentation strategy to the segmented CSI samples during training. The purpose is to expose the model to deployment variations that are common in practical homes but underrepresented in a limited training set. Unlike generic image augmentations such as rotation or flipping, or artificial frame-insertion methods used in recent CSI studies \cite{b30}, the following operations are explicitly chosen to remain consistent with wireless propagation characteristics.
	
	\subsubsection{Rician Noise Injection}
	Indoor WiFi propagation is affected by thermal noise, hardware imperfections, and small-scale fading. To mimic these perturbations, we inject additive Gaussian noise
	\begin{equation}
		N \sim \mathcal{N}(0,\sigma^2)
	\end{equation}
	into the CSI sample during training. This discourages the model from overfitting to environment-specific noise realizations.
	
	\subsubsection{Random Amplitude Scaling}
	The received CSI amplitude varies with deployment distance and path loss.
	A model trained only on one room scale may incorrectly associate strong signal magnitude with fall events.
	To remove this shortcut and simulate independent channel fading variations, we randomly scale the CSI amplitude independently for each receive antenna channel:
	\begin{equation}
		X'_{scale}(c,:,:) = \lambda_c \cdot X(c,:,:), \quad \lambda_c \sim \mathcal{U}(0.5, 1.5),
	\end{equation}
	where $\lambda_c$ is an independent scaling factor drawn for each of the $c \in \{1, 2, 3\}$ channels.
	This encourages the network to rely on motion morphology rather than absolute or relative antenna signal intensity.
	
	\subsubsection{Random Time Shift}
	In real-world monitoring, a fall may occur at any temporal position within a detection window. To avoid overfitting to a fixed event location, we circularly shift the sample along the time axis:
	\begin{equation}
		X'_{shift}(t) = X((t-\delta)\bmod T), \quad \delta \in [-50,50].
	\end{equation}
	This improves temporal invariance and prevents the temporal encoder from memorizing a narrow event onset pattern.
	
	\subsubsection{NLoS Simulation via Spectral Smoothing}
	One of the most challenging deployment conditions is NLoS propagation, in which obstacles attenuate high-frequency motion components and blur the CSI response of a fall. To simulate this effect during training, we probabilistically degrade a sample by down-sampling and interpolation:
	\begin{equation}
		X_{down} = \text{DownSample}(X, \text{scale}=0.5),
	\end{equation}
	\begin{equation}
		X_{NLoS} = \text{Interp}(X_{down}, \text{size}=T).
	\end{equation}
	This operation preserves the low-frequency motion envelope while suppressing sharp temporal details, thereby approximating the spectral smoothing effect of NLoS environments.
	
	\subsection{Channel Standardization and Dynamic Variance Gate}
	Before deep feature extraction, the augmented instance-normalized tensor undergoes a final ImageNet-style channel standardization (using means $[0.485, 0.456, 0.406]$ and standard deviations $[0.229, 0.224, 0.225]$). This aligns the data distribution with the pre-trained EfficientNet backbone's expected input space.
	
	Even after standardization, CSI samples still contain strong static environmental components arising from walls, furniture, and stable multipath reflections. In contrast, human motion manifests as local temporal fluctuations superimposed on this background. To explicitly separate these two factors, we introduce a DVG at the front of the network. The key idea is that motion-sensitive regions should exhibit larger short-term variance than static clutter. 
	
	Given the standardized tensor $X'' \in \mathbb{R}^{C \times T \times S}$, we compute local temporal variance using a 1D sliding window of size $W=15$ strictly along the temporal axis. To maintain the temporal dimension $T$, replicate padding is applied to the sequence edges before average pooling. Using the identity $Var(Z)=E[Z^2]-(E[Z])^2$, the variance map is efficiently implemented:
	\begin{equation}
		V_{t,s} = \text{ReLU}\left(E[X''^2]_W - (E[X'']_W)^2\right) + \epsilon,
	\end{equation}
	where $\epsilon = 10^{-6}$.
	
	To convert this physically interpretable variance map into a learnable gating mask, we scale the variance to amplify motion signals and apply a $3 \times 3$ convolution followed by a sigmoid activation:
	\begin{equation}
		M_{var} = \sigma(\text{Conv}_{3 \times 3}(\alpha \cdot V)),
	\end{equation}
	where $\alpha$ is a scaling factor. Since normalized variance values are typically infinitesimal, this explicit scaling provides sufficient gradient momentum to overcome the negative bias constraint during high-intensity dynamic motions. 
	
	To embed a strict physical prior into the network, we explicitly initialize the bias of the convolutional layer to a negative constant ($-3.0$). This ensures that in the absence of motion (where variance approaches zero), the sigmoid activation natively suppresses the static background to near zero ($\sigma(-3.0) \approx 0.047$), acting as a robust environmental filter from the very first training epoch. The convolutional weights are correspondingly initialized with a mean of $1/9$ to act as a spatial smoothing filter. The final gated representation is then obtained by $X_{gate} = X'' \otimes M_{var}$. 
	
	Ultimately, the DVG serves as the core physics-driven ``environment decoupler'' of the proposed framework, fundamentally resolving the cross-domain generalization bottleneck across four dimensions. At the \textit{physical level}, it acts as an environmental eraser: by leveraging the negative bias to suppress zero-variance DC components, it strips away the room-specific multipath reflections caused by static objects. At the \textit{signal level}, it isolates and amplifies the AC components: human activities induce sharp localized variance fluctuations, which the DVG instantaneously highlights via the scaling factor $\alpha$, thereby pinpointing the pure dynamic motion. At the \textit{system level}, this early-stage filtering prevents the CNN backbone from memorizing the background layout of the training environment, effectively paving the way for true zero-shot cross-domain transfer to unseen deployment sites. Finally, at the \textit{architectural level}, DVG offloads the heavy burden of noise suppression from the backbone. By providing a clean, motion-isolated representation, it allows the subsequent CBAM and Transformer modules to focus exclusively on modeling the invariant spatiotemporal semantics of the fall event, preventing overfitting and ensuring a physics-aware learning process.
	
	\subsection{Attention-Enhanced CNN-Transformer Backbone}
	After preprocessing and DVG-based gating, the CSI tensor is passed to a hybrid backbone that combines convolutional encoding, attention-based feature refinement, and Transformer-based temporal reasoning. This design reflects the structure of a fall event itself: local time-subcarrier patterns must first be extracted, then irrelevant background responses should be suppressed, and finally the full temporal evolution of the event should be modeled over a longer horizon.
	
	\subsubsection{Local Feature Extraction with EfficientNet-B0}
	To extract local spatiotemporal features from the gated tensor $X_{gate} \in \mathbb{R}^{3 \times 625 \times 30}$, we adopt the EfficientNet-B0 architecture originally adapted for WiFi CSI by Chu \textit{et al.} \cite{b14}. The standard EfficientNet is designed to classify images in the ImageNet dataset, which consists of 1000 classes. To tailor the network for WiFi-based sensing, we treat the gated CSI tensor as an image-style input.
	
	As detailed in Table \ref{tab:efficientnet_backbone}, the shared architectural layers of the original EfficientNet-B0 are retained up to Stage 9, mapping the input to high-level feature maps
	\begin{equation}
		F \in \mathbb{R}^{C \times H \times W},
	\end{equation}
	where $C=1280$, $H=20$, and $W=1$ for the final backbone output. Unlike the modified architecture in \cite{b14} which applies global pooling and fully connected layers directly for binary classification, we discard the original top structure. Instead, the extracted spatiotemporal feature map $F$ is preserved in its spatial and temporal dimensions, serving as the foundational input for the subsequent attention refinement and sequence modeling stages.
	
	Furthermore, transfer learning is utilized by initializing the backbone with ImageNet pre-trained weights. This leverages robust baseline feature representations and accelerates network convergence. Overall, this adapted EfficientNet provides an optimal trade-off between representational capacity and computational efficiency, making it highly suitable as the front-end feature extractor in our edge-oriented framework.
	
	\begin{table}[htbp]
		\centering
		\caption{Architecture of the EfficientNet-B0 Feature Extractor}
		\label{tab:efficientnet_backbone}
		\renewcommand{\arraystretch}{1.6}
		\begin{tabular}{|c|c|c|}
			\hline
			\textbf{Stage} & \textbf{Operator / Processing} & \textbf{Output Size} \\
			\hline
			Input & Gated CSI Tensor ($X_{gate}$) & $3 \times 625 \times 30$ \\
			\hline
			Stage 1 & Conv $3 \times 3$, stride 2 & $32 \times 313 \times 15$ \\
			\hline
			Stage 2 & $\begin{bmatrix} \text{MBConv1, k}3\times3 \end{bmatrix} \times 1$ & $16 \times 313 \times 15$ \\
			\hline
			Stage 3 & $\begin{bmatrix} \text{MBConv6, k}3\times3 \end{bmatrix} \times 2$ & $24 \times 157 \times 8$ \\
			\hline
			Stage 4 & $\begin{bmatrix} \text{MBConv6, k}5\times5 \end{bmatrix} \times 2$ & $40 \times 79 \times 4$ \\
			\hline
			Stage 5 & $\begin{bmatrix} \text{MBConv6, k}3\times3 \end{bmatrix} \times 3$ & $80 \times 40 \times 2$ \\
			\hline
			Stage 6 & $\begin{bmatrix} \text{MBConv6, k}5\times5 \end{bmatrix} \times 3$ & $112 \times 40 \times 2$ \\
			\hline
			Stage 7 & $\begin{bmatrix} \text{MBConv6, k}5\times5 \end{bmatrix} \times 4$ & $192 \times 20 \times 1$ \\
			\hline
			Stage 8 & $\begin{bmatrix} \text{MBConv6, k}3\times3 \end{bmatrix} \times 1$ & $320 \times 20 \times 1$ \\
			\hline
			Stage 9 & Conv $1 \times 1$ & $1280 \times 20 \times 1$ \\
			\hline
		\end{tabular}
	\end{table}
	\subsubsection{Feature Refinement with CBAM}
	Although the gated tensor already suppresses much of the static clutter, residual irrelevant responses may still remain across channels and spatial locations. To further refine the feature maps, we insert a CBAM after the backbone. CBAM sequentially performs channel attention and spatial attention.
	
	For channel attention, average pooling and max pooling are applied across the spatial dimensions, and the resulting descriptors are passed through a shared multilayer perceptron:
	\begin{equation}
		M_c(F) = \sigma(MLP(\text{AvgPool}(F)) + MLP(\text{MaxPool}(F))).
	\end{equation}
	The backbone features are then reweighted as
	\begin{equation}
		F' = M_c(F) \otimes F.
	\end{equation}
	This step emphasizes informative feature channels and suppresses noisy ones.
	
	For spatial attention, average pooling and max pooling are applied across the channel dimension, concatenated, and followed by a $7 \times 7$ convolution:
	\begin{equation}
		M_s(F') = \sigma\left(f^{7 \times 7}([\text{AvgPool}(F');\text{MaxPool}(F')])\right).
	\end{equation}
	The refined feature map is obtained as
	\begin{equation}
		F'' = M_s(F') \otimes F'.
	\end{equation}
	This operation helps the model focus on the most informative time-subcarrier regions associated with the fall event.
	
	\subsubsection{Temporal Modeling with Transformer Encoder}
	Although the features refined by CBAM have been stripped of environmental noise and subcarrier-level interference, they remain in a high-dimensional feature space. To interpret these features as a cohesive temporal event, we transition from spatial-channel analysis to sequential reasoning using a Transformer encoder. 
	
	First, we employ a $1 \times 1$ convolutional layer as a linear projection to reduce the channel dimension from 1280 to $d_{model}=512$. To focus strictly on the temporal evolution, we collapse the frequency (subcarrier) dimension via mean pooling, resulting in a temporal embedding sequence $E \in \mathbb{R}^{T' \times d_{model}}$. This sequence represents the purified motion energy across time, now decoupled from room-specific multipath reflections thanks to the preceding DVG module.
	
	Given that the self-attention mechanism is inherently permutation-invariant, it cannot naturally perceive the sequential order of a fall—a process where the timing of the transition from rapid descent to impact is critical. Therefore, we inject sinusoidal Positional Encodings into the sequence to preserve the chronological context of the fall stages. 
	
	The position-aware sequence is then processed by a Transformer encoder consisting of $L=2$ layers, each featuring $h=4$ attention heads. The core multi-head self-attention (MHSA) mechanism \cite{b23} allows the model to relate distant time steps, such as the pre-fall instability and the post-fall stillness, within a single unified context:
	\begin{equation}
		\text{Attention}(Q,K,V)=\text{softmax}\left(\frac{QK^T}{\sqrt{d_k}}\right)V,
	\end{equation}
	where $Q, K, V$ are derived from the same refined temporal embeddings. By attending to these dependencies, the Transformer can distinguish a genuine fall from confounding activities like sitting down, which may share similar local variances but exhibit different long-term temporal signatures. 
	
	Finally, the temporal sequence is aggregated via global average pooling, followed by an MLP classifier with SiLU activation and dropout ($p=0.3$) to output the final detection probability. This hierarchical integration of physics-driven gating, spatial-channel refinement, and Transformer-based temporal reasoning ensures the system's robust performance in complex, unseen environments.

	\subsection{Loss Function: Focal Loss}
	Fall detection datasets are inherently imbalanced, and the most safety-critical samples, such as weak NLoS falls, are often also the hardest to classify. To address both class imbalance and hard-example emphasis, we adopt focal loss:
	\begin{equation}
		L_{FL} = -\alpha(1-p_t)^\gamma \log(p_t),
	\end{equation}
	where $p_t$ is the predicted probability of the ground-truth class, $\alpha$ is the class-balancing factor, and $\gamma$ is the focusing parameter. In our implementation, $\gamma=2.0$ is used to down-weight easy samples and concentrate learning on ambiguous or low-quality fall events, while $\alpha=3.0$ increases the penalty on missed fall detections. This objective is well aligned with the practical requirement of prioritizing recall in elderly safety monitoring.

	\section{Experiments}
	
	\subsection{Experimental Setup}
	\subsubsection{Dataset}
\begin{figure}[t]
		\centering
		\includegraphics[width=0.9\columnwidth]{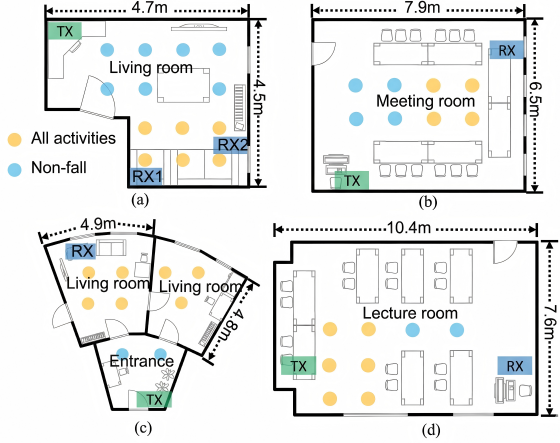}
		\caption{Dataset Environments}
		\vspace{-6pt}
		\label{sm}
	\end{figure}

	We utilize a comprehensive dataset \cite{b14} collected using Intel 5300 NICs across four distinct indoor environments, as shown in Fig \ref{sm}: a Living room, a Meeting room, a Home lab (divided into LoS and NLoS areas), and a Lecture room. To ensure a high degree of diversity in human movement, this dataset recruited 22 volunteers (aged 24 to 43, 7 women and 15 men) to participate in various activities, ultimately recording 321 fall events and 436 non-fall daily activities which is shown in Table \ref{tab:events_collected}.The fall events cover four directions (front, back, left, and right) and varying impact intensities, preceded by random movements such as standing still, walking forwards, or moving backwards. The non-fall activities include bending to pick up an object, sitting down, standing up, walking, and waving arms. Furthermore, these activities were executed on both dominant LoS paths and NLoS paths to capture realistic signal propagation complexity. To rigorously evaluate domain generalization, we adopt a ``Leave-One-Environment-Out'' cross-validation strategy, where the training and testing sets always belong to disjoint environments.
	
	\begin{table}[htbp]
		\caption{Number of events collected in the dataset.}
		\label{tab:events_collected}
		\centering
		\begin{tabular}{lccc}
			\toprule
			\multicolumn{1}{c}{\textbf{Environments}} & \textbf{Fall events} & \textbf{Non-fall events} & \textbf{Total} \\
			\midrule
			Living room A       & 91 & 157 & 248 \\
			Meeting room B      & 58 & 76  & 134 \\
			Left living room C  & 40 & 45  & 85  \\
			Right living room C & 36 & 49  & 85  \\
			Lecture room D      & 96 & 109 & 205 \\
			\bottomrule
		\end{tabular}
	\end{table}
	\subsubsection{Implementation Details}
	The model is implemented in PyTorch and trained on an NVIDIA GPU H100. We use the Adam optimizer with an initial learning rate of $5 \times 10^{-4}$ and a cosine annealing schedule. During inference, we employ test-time augmentation (TTA), averaging predictions over 5 slightly perturbed versions of the input to further stabilize performance.
	
	\subsection{Performance Analysis}
	\subsubsection{Scenario 1: Overall Baseline Performance on the Aggregated Dataset}
	To establish a foundational baseline for our proposed architecture before evaluating its cross-domain generalization capabilities, we first conduct an experiment using the complete aggregated dataset. In this scenario, all data samples from the four distinct environments (Living Room A, Meeting Room B, Left/Right Living Room C, and Lecture Room D) were combined into a single unified dataset. We applied a standard random split—allocating 80 \% of the data for training and 20 \% for testing. This ensures that the model is exposed to the static background noise, spatial layouts, and both LoS and NLoS propagation paths of all rooms during the training phase.
	
	Since the training and testing sets share identical environmental distributions in this setting, this scenario primarily assesses the pure feature extraction and spatiotemporal representation capacities of the models under ideal, in-domain conditions.As shown in Table \ref{scenario_baseline}, under these mixed-environment settings, our proposed Attention-Enhanced CNN-Transformer framework demonstrates exceptional representational power, achieving near-perfect classification accuracy and consistently outperforming baseline methods such as FallDeFi and the standard CNN-based approaches.
	
	The superior performance in this baseline scenario validates the efficacy of our core architectural design: the integration of the CBAM module effectively refines local time-subcarrier patterns by adaptively weighting essential spatial-channel features, while the Transformer Encoder perfectly captures the causal, multi-stage temporal evolution of fall events. This confirms that when sufficient environmental context is provided, our hybrid architecture possesses the requisite capacity to accurately distinguish subtle fall morphologies from complex daily confounding activities.
	\setlength{\tabcolsep}{4mm}
	\begin{table}[htbp]
		\caption{Performance comparison of classifiers trained and tested on the aggregated dataset.}
		\label{scenario_baseline}
		\centering
		\begin{tabular}{lccc}
			\toprule
			\multirow{2}{*}{\textbf{Model}} & \multicolumn{3}{c}{\textbf{Aggregated Dataset}} \\
			\cmidrule(lr){2-4}
			& \textbf{Acc} & \textbf{Prec} & \textbf{Rec} \\
			\midrule
			Modified B0 \cite{b14}   & 94.9\% & 91.7\% & 96.8\% \\
			FallDeFi                 & 80.8\% & 75.0\% & 81.9\%  \\
			Work in \cite{b4}        & 88.0\% & 82.9\% & 90.2\%  \\
			\textbf{Proposed}        & \textbf{99.5\%} & \textbf{98.8\%} & \textbf{100.0\%} \\
			\bottomrule
		\end{tabular}
	\end{table}
	\subsubsection{Scenario 2: Robustness to Challenging NLoS Environments}
	Building upon the overall performance evaluation in Scenario 1, we further investigate the model's reliability under the most demanding conditions within the mixed-environment deployment: the NLoS scenario. In this experiment, while the training set remains the entire aggregated dataset (ensuring the model is exposed to maximal environmental diversity), we specifically isolate and evaluate performance on the NLoS test subset (Right Living Room C). This enables us to assess whether the architecture can effectively distinguish fall morphologies when the signal is severely compromised by physical obstructions.
	
	As summarized in Table \ref{tab:scenario1}, conventional models exhibit a significant performance bottleneck in this scenario despite being trained on the full dataset. Architectures such as FallDeFi \cite{b15} and the CNN-based approach in \cite{b4} suffer a notable decline in accuracy, both dropping below 76\%. This performance degradation is primarily driven by two compounding factors. First, the inherent data imbalance—where NLoS samples constitute only a small fraction (85 out of 757 total events)—limits the ability of standard classifiers to learn the subtle distribution of obstructed signals. More fundamentally, physical barriers act as environment-induced low-pass filters that attenuate the high-frequency Doppler components essential for fall recognition. The resulting smoothed NLoS fall signatures closely mimic the CSI patterns of benign, low-energy activities (e.g., sitting or walking) executed in LoS environments, leading to severe confusion in models that lack explicit motion-enhancement mechanisms.
	
	In stark contrast, our proposed framework demonstrates exceptional resilience, sustaining an accuracy of \textbf{97.6\%}. This superior performance validates the efficacy of the DVG in decoupling environment-specific DC components from motion-induced AC fluctuations. By adaptively amplifying local temporal variances, our model successfully extracts invariant fall signatures even from spectrally smoothed NLoS streams, effectively bridging the safety coverage gap in obstructed home environments.

	\setlength{\tabcolsep}{4mm}
	\begin{table}[t]
		\caption{Performance comparison of classifiers trained with the entire dataset and tested in the NLoS dataset.}
		\label{tab:scenario1}
		\centering
		\begin{tabular}{lccc}
			\toprule
			\multirow{2}{*}{\textbf{Model}} & \multicolumn{3}{c}{\textbf{Right Living Room C}} \\
			\cmidrule(lr){2-4}
			& \textbf{Acc} & \textbf{Prec} & \textbf{Rec} \\
			\midrule
			Modified B0 \cite{b14}   & 82.0\% & 72.7\% & 100.0\% \\
			FallDeFi & 68.2\% & 66.4\% & 70.9\%  \\
			Work in \cite{b4} & 75.4\% & 72.7\% & 76.2\%  \\
			\textbf{Proposed} & \textbf{97.6\%} & \textbf{97.3\%} & \textbf{100.0\%} \\
			\bottomrule
		\end{tabular}
	\end{table}
	\subsubsection{Scenario 3: Generalization to Unseen Room A and D}
\begin{figure}[htbp]
		\centering
		\includegraphics[width=0.9\columnwidth]{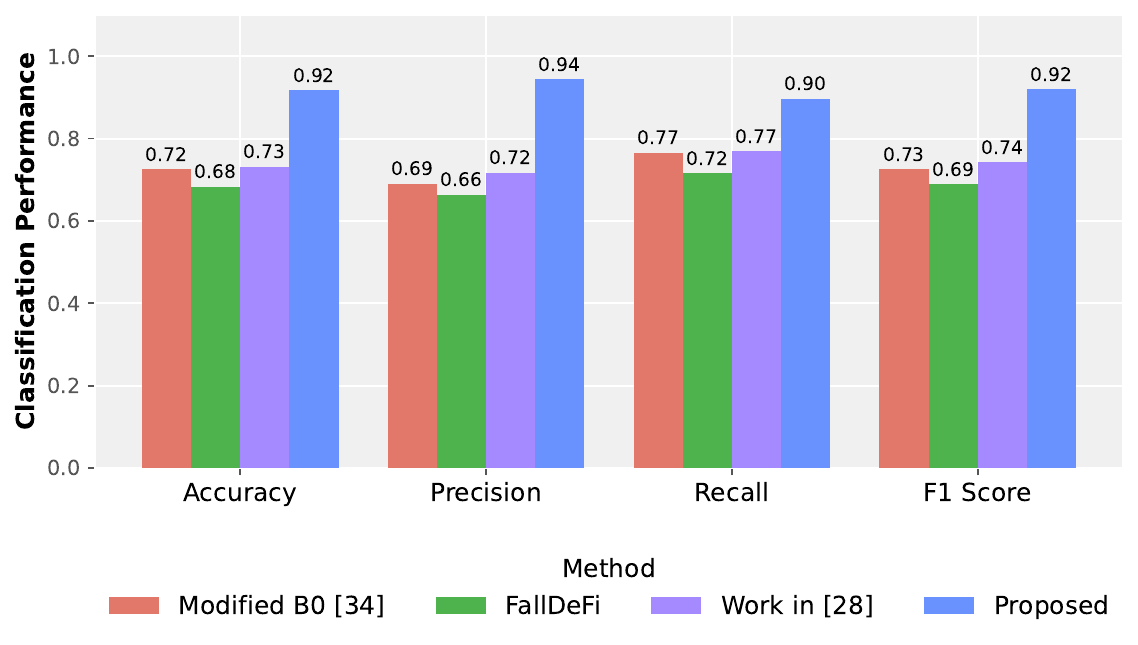}
		\caption{Classification Performance in Lecture Room D (Scenario 3).}
		\label{fig:scenario1_lecture_d}
		\vspace{0.2cm} 
		\includegraphics[width=0.9\columnwidth]{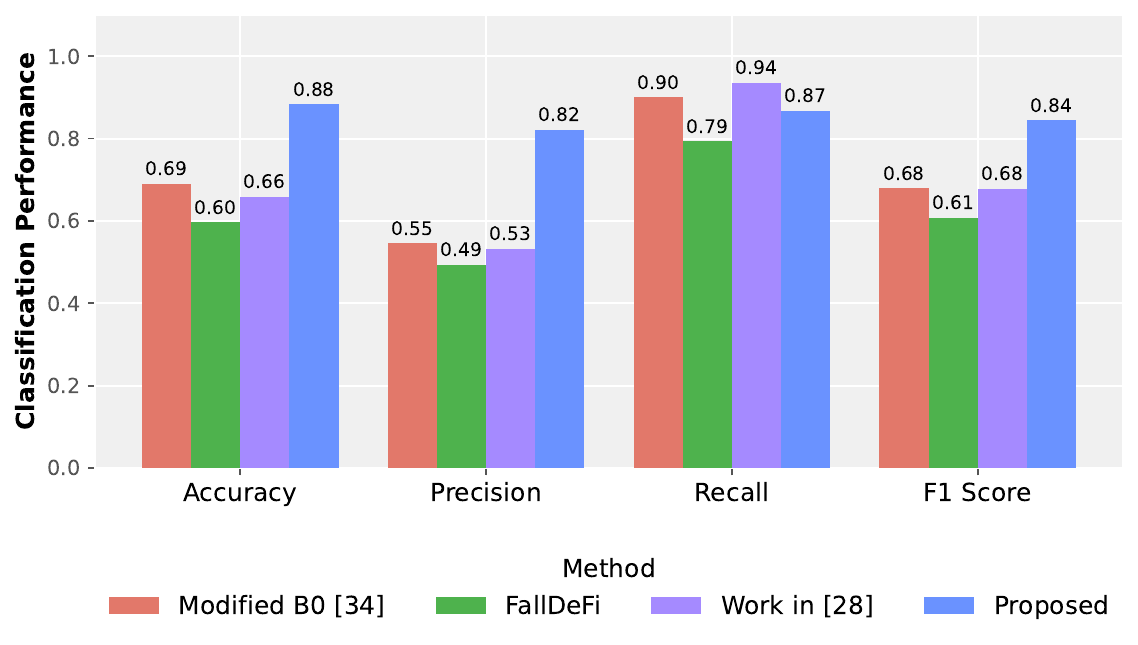}
		\caption{Classification Performance in Living Room A (Scenario 3).}
		\label{fig:scenario1_living_a}
	\end{figure}

	In  Fig \ref{fig:scenario1_lecture_d} and Fig \ref{fig:scenario1_living_a}, we evaluate a severe domain shift scenario by training the model on Room B and Room C (both LoS and NLoS areas) and testing it on the completely unseen Living Room A and Lecture Room D. The differing room dimensions and furniture layouts cause the baseline EfficientNet-B0 accuracy to plummet to 69.0\% and 72.5\%. 
	
	In stark contrast, our proposed method maintains a robust accuracy of \textbf{88.3\%} and \textbf{91.7\%}. This superior generalization is largely attributed to the integration of the CBAM module. Specifically, the channel attention mechanism adaptively suppresses environment-specific background noise across irrelevant subcarriers, while the spatial attention forces the network to focus on the invariant spatiotemporal signature of the fall. Consequently, the Transformer Encoder processes highly refined, domain-agnostic features, enabling reliable temporal modeling even in a novel environment.
	
	\subsubsection{Scenario 4: Sparse Training Data}
\begin{figure}[htbp]
		\centering
		\includegraphics[width=0.9\columnwidth]{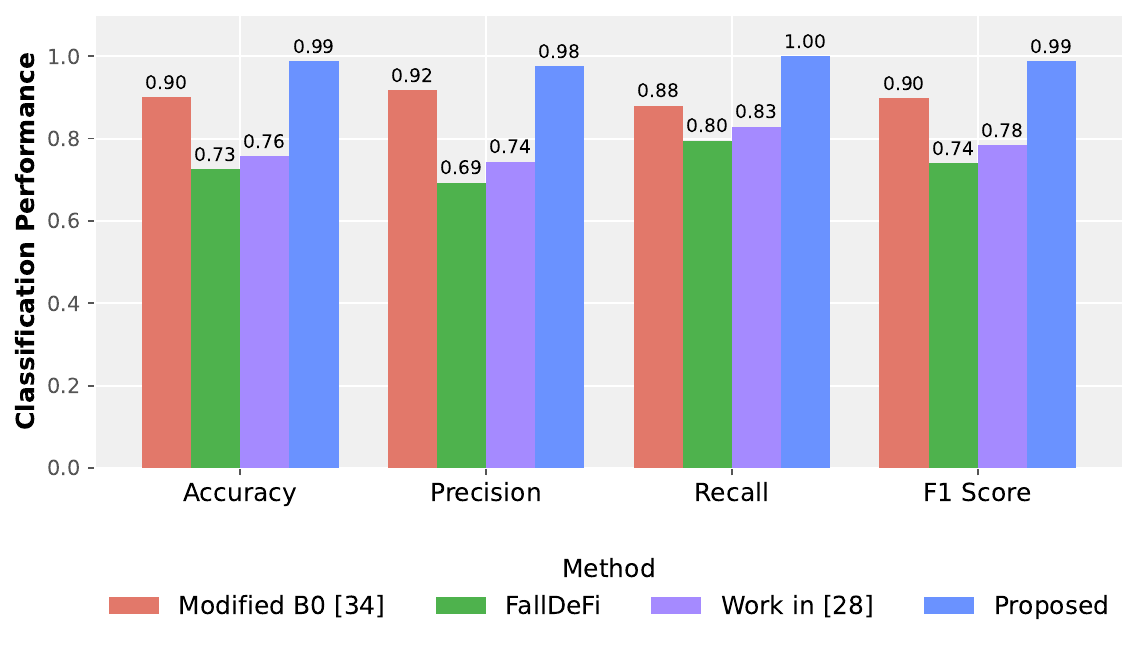}
		\caption{Classification Performance in Left Living Room C (Scenario 4).}
		\label{fig:scenario1_cl}
		\vspace{0.2cm} 
		\includegraphics[width=0.9\columnwidth]{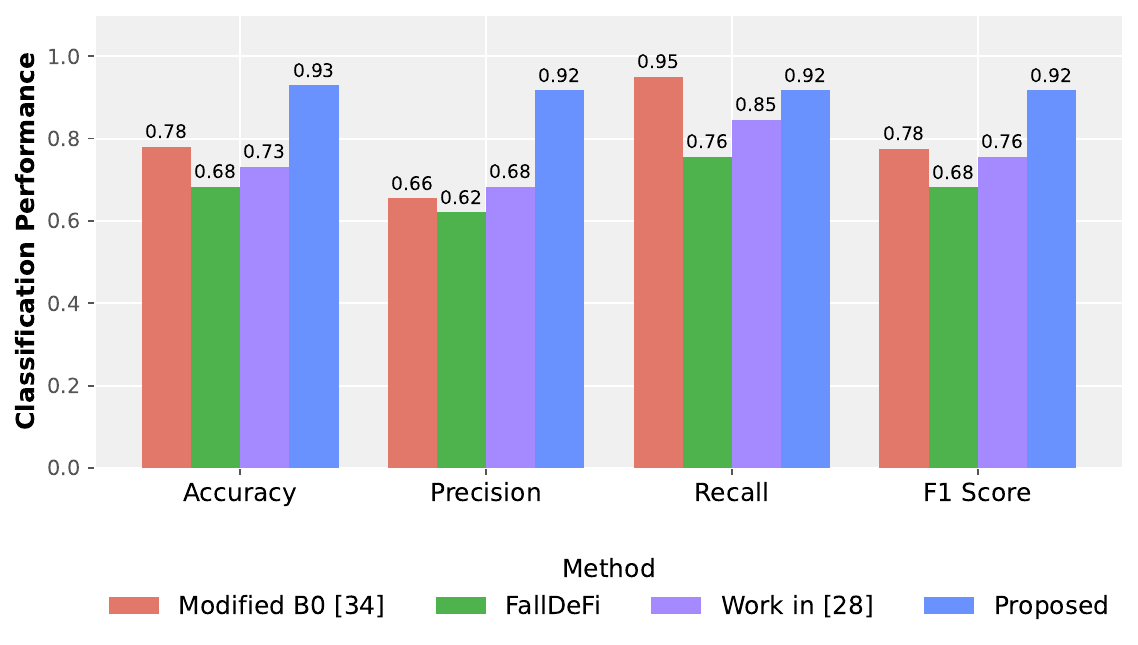}
		\caption{Classification Performance in Right Living Room C (Scenario 4).}
		\label{fig:scenario1_cr}
	\end{figure}

	Fig \ref{fig:scenario1_cl} and Fig \ref{fig:scenario1_cr} evaluates the model's robustness when trained on a sparse and disjoint data distribution (restricted to only Meeting Room B and Lecture Room D). Testing is conducted on both the LoS (Left) and NLoS (Right) areas of Home Lab C. Even with limited environmental diversity during training, our method achieves \textbf{98.8\%} accuracy in the LoS setting and an impressive \textbf{92.9\%} in the challenging NLoS setting. 
	
	In this sparse-data scenario, the dual-attention mechanism of CBAM acts as a vital feature recalibrator. It prevents the CNN backbone from overfitting to the static background clutter of the limited training rooms. By extracting the core motion features through attention maps before sequence modeling, our hybrid architecture vastly outperforms existing schemes like FallDeFi (68.3\%) and the standard CNN-based approach in \cite{b4} (73.2\%) by a wide margin.
	
	\subsubsection{Ablation Study}
	To rigorously evaluate the individual contribution of each proposed module, we conduct a comprehensive ablation study.
	The standard EfficientNet-B0 (pre-trained on ImageNet) with global normalization serves as our baseline.
	We then incrementally integrate our proposed components: Instance Normalization combined with Physics-Aware Data Augmentation (IN+PA), the novel DVG, the CBAM, and finally, the Transformer Encoder.
	Evaluations are performed on two highly challenging cross-domain test sets corresponding to our previous scenarios: the completely unseen Living Room A (adopting the setup from Scenario 3, trained on Rooms B and C to assess severe spatial domain shift) and the Right Living Room C (adopting the sparse-data setup from Scenario 4, trained on Rooms B and D to assess NLoS signal attenuation).
	As detailed in Table \ref{tab:ablation}, the baseline model struggles significantly in these unseen and obstructed environments, achieving only 69.0\% and 78.0\% accuracy, respectively.
	The introduction of IN and PA-DA provides an immediate performance boost by eliminating absolute amplitude dependence and forcing the network to learn low-frequency precursors robust to NLoS conditions.
	Crucially, the integration of the proposed DVG module yields a substantial improvement in both scenarios (e.g., boosting unseen Room A accuracy from 78.1\% to 82.4\%).
	This validates our physical hypothesis: by dynamically calculating local temporal variance, DVG effectively filters out the static environmental DC components (reflections from unseen walls and furniture) and cleanly decouples the dynamic human motion signatures.
	Furthermore, the addition of CBAM adaptively refines the spatiotemporal feature maps by suppressing residual subcarrier noise.
	Finally, integrating the Transformer Encoder optimally models the long-range temporal dynamics of the fall process, culminating in the highest overall accuracies of 88.3\% and 92.9\% across the respective test scenarios.
	
	\begin{table}[htbp]
		\caption{Ablation Study on Key Components of the Proposed Framework}
		\begin{center}
			\setlength{\tabcolsep}{3pt}
			\begin{tabular}{lcccccc}
				\toprule
				\multirow{2}{*}{\textbf{Model Variant}} & \multicolumn{4}{c}{\textbf{Components}} & \multicolumn{2}{c}{\textbf{Accuracy (\%)}} \\
				\cmidrule(lr){2-5} \cmidrule(lr){6-7}
				& \textbf{IN+PA} & \textbf{DVG} & \textbf{CBAM} & \textbf{Trans.} & \textbf{Room A} & \textbf{Right Room C} \\
				\midrule
				Baseline  & & & & & 69.0 & 78.0 \\
				Variant 1 & $\checkmark$ & & & & 78.1 & 83.5 \\
				Variant 2 & $\checkmark$ & $\checkmark$ & & & 82.4 & 88.2 \\
				Variant 3 & $\checkmark$ & $\checkmark$ & $\checkmark$ & & 85.6 & 90.6 \\
				\textbf{Proposed} & $\checkmark$ & $\checkmark$ & $\checkmark$ & $\checkmark$ & \textbf{88.3} & \textbf{92.9} \\
				\bottomrule
			\end{tabular}
			\label{tab:ablation}
		\end{center}
	\end{table}
	
	\subsection{Real-Time Field Testing and Edge Deployment}

	\subsubsection{Live Inference System Setup}
	To validate the practical utility of our proposed framework for IoT-enabled healthcare, we extend our evaluation from offline datasets to a real-time live inference system. 
	The edge sensing platform comprises two host computers equipped with commodity Intel 5300 NICs. 
	One computer operates in injection mode, acting as a WiFi Access Point that broadcasts CSI packets at a rate of 1000 packets/s. 
	The secondary PC, acting as the edge computing node, operates in monitor mode, passively acquiring the real-time CSI data stream without needing to join a specific WiFi network.
	\begin{figure*}[htbp]
		\centering
		\includegraphics[width=\textwidth]{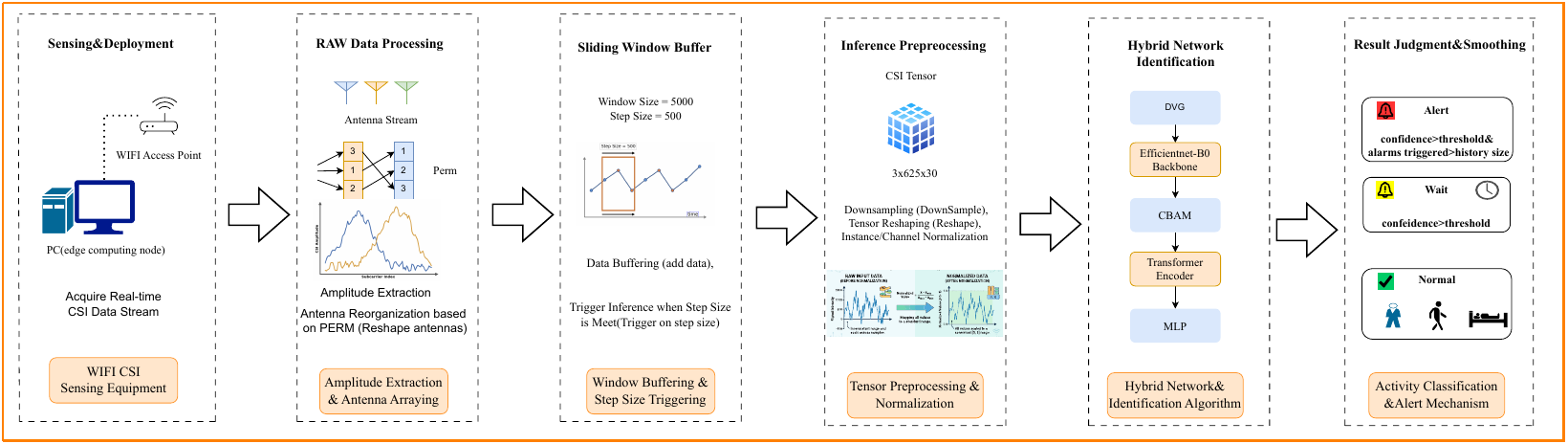}
		\caption{Real-time testing flowchart.}
		\label{fig:real}
	\end{figure*}
	As illustrated in Fig \ref{fig:real}, the live inference pipeline executes a sequence of strictly synchronized operations:
	\begin{itemize}
		\item RAW Data Processing: The system continuously reads incoming packets to extract the raw amplitude. Crucially, it performs antenna reorganization based on the dynamic permutation array (PERM) to ensure the physical consistency of the spatial streams before amplitude extraction and antenna arraying.
	
		\item Sliding Window Buffer: To capture the complete temporal morphology of a fall while maintaining responsiveness, the processed stream flows into a buffer with a window size of 5000 packets (equivalent to 5 seconds). A step size of 500 packets (0.5 seconds) is utilized to trigger the downstream inference, ensuring an overlapping data buffering that prevents the truncation of sudden events.
	
		\item Inference Preprocessing: Upon triggering, the buffered data undergoes a 4th-order Butterworth low-pass filtering and temporal down-sampling by a factor of 8, reducing the temporal dimension to 625 points. The data is subsequently reshaped into a $3 \times 625 \times 30$ CSI Tensor, followed by instance and channel normalization to mitigate environment-specific amplitude bias.
	
		\item Hybrid Network Identification: This standardized tensor is then fed into our Hybrid Network Identification Algorithm. The physical-driven DVG filters out static environmental reflections. The signal is then processed by the EfficientNet-B0 backbone, refined by the CBAM module, and evaluated by the Transformer Encoder before the final MLP outputs the temporal classification probability.
	
		\item Result Judgment and Smoothing: Finally, to minimize false alarms in edge deployments, we implement an activity classification and alert mechanism using a temporal smoother. If the inference confidence exceeds the predefined threshold (e.g., 50\%) for a single frame, the system flags a ``Wait'' status  (suspected fall). A confirmed ``Alert'' is strictly triggered only when consecutive alarms exceed the history size. Otherwise, it outputs a ``Normal'' state. This logic provides a robust, low-false-positive monitoring solution suitable for real-world elderly care.
	\end{itemize}

	\subsubsection{Cross-Domain Field Robustness}
\begin{figure}[t]
		\centering
		\includegraphics[width=0.8\columnwidth]{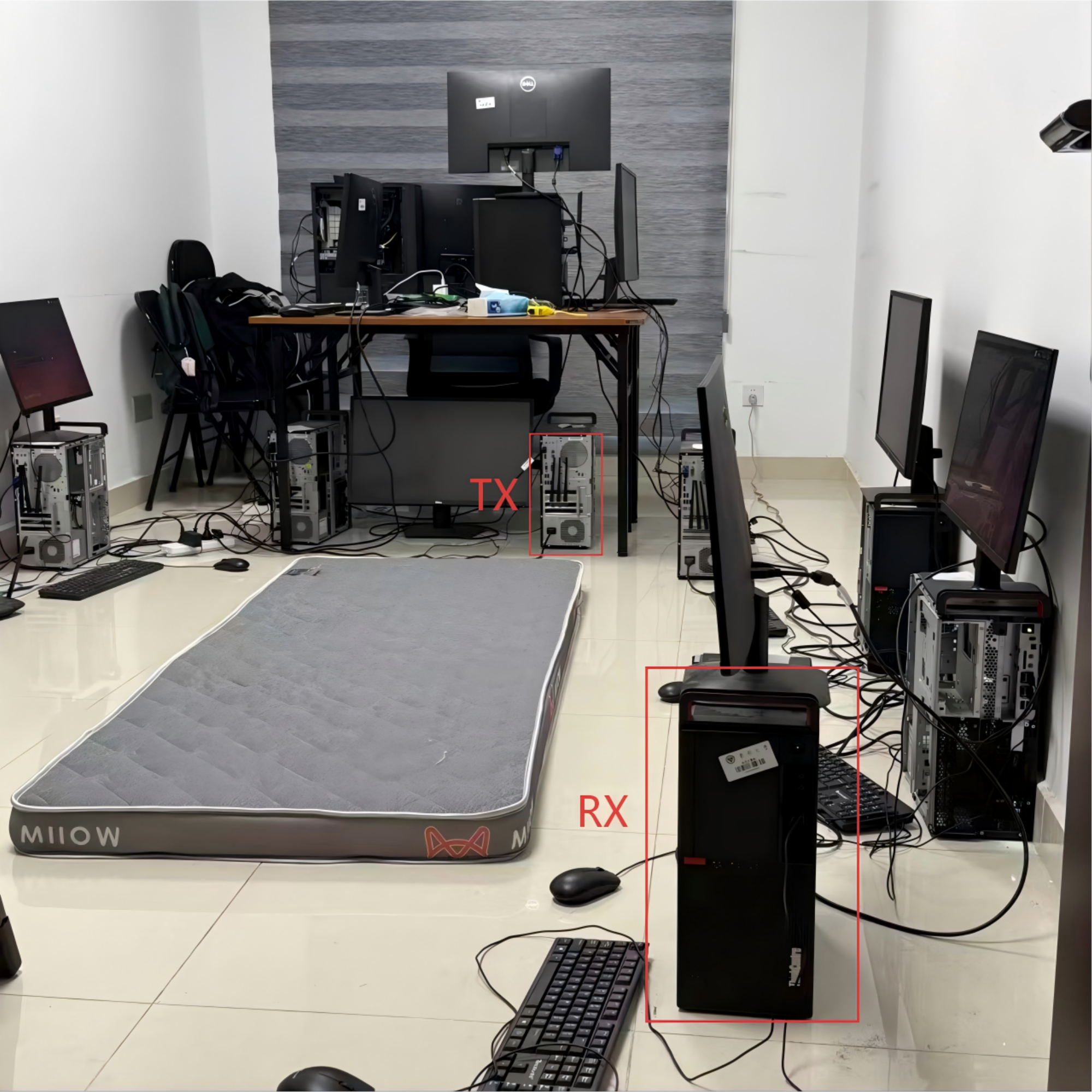}
		\caption{Field Testing Environment}
		\label{room}
	\end{figure}

\begin{figure}[!t]
		\centering
		\subfloat[]{\includegraphics[width=0.4\linewidth]{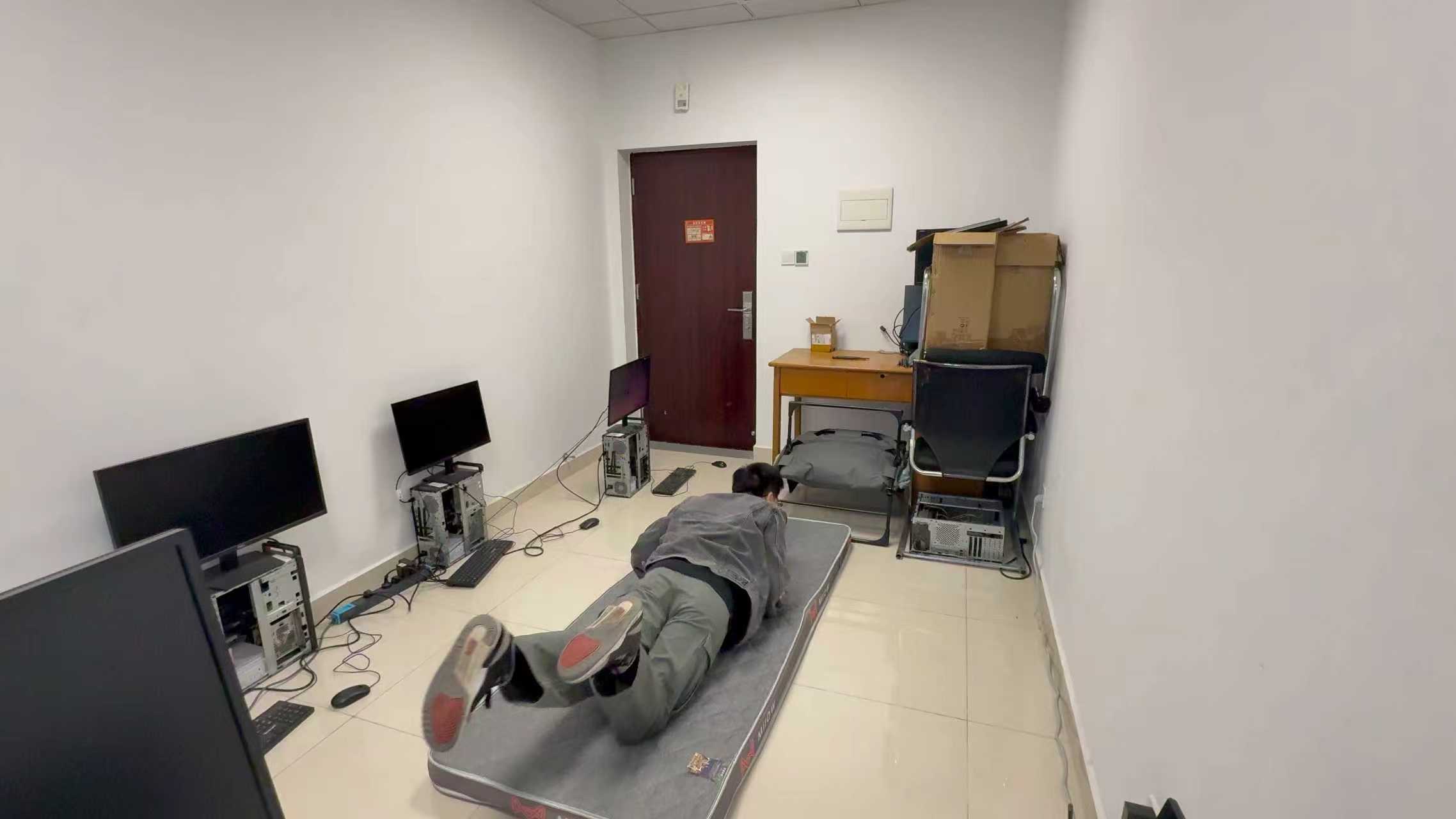}
			\label{fig_first_case}}
		\hfil 
		\subfloat[]{\includegraphics[width=0.4\linewidth]{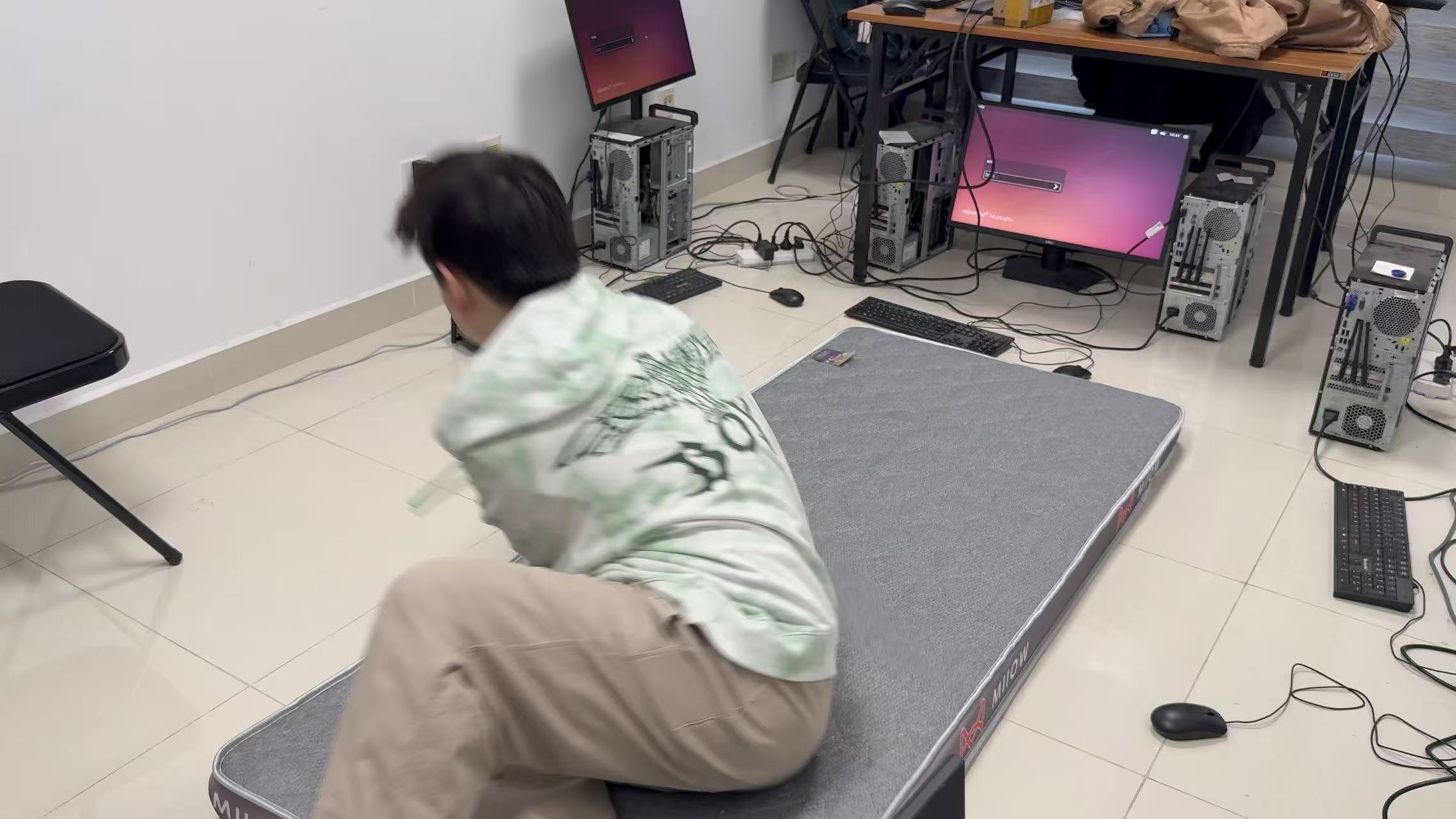}
			\label{fig_second_case}}
		\hfil
		\subfloat[]{\includegraphics[width=0.4\linewidth]{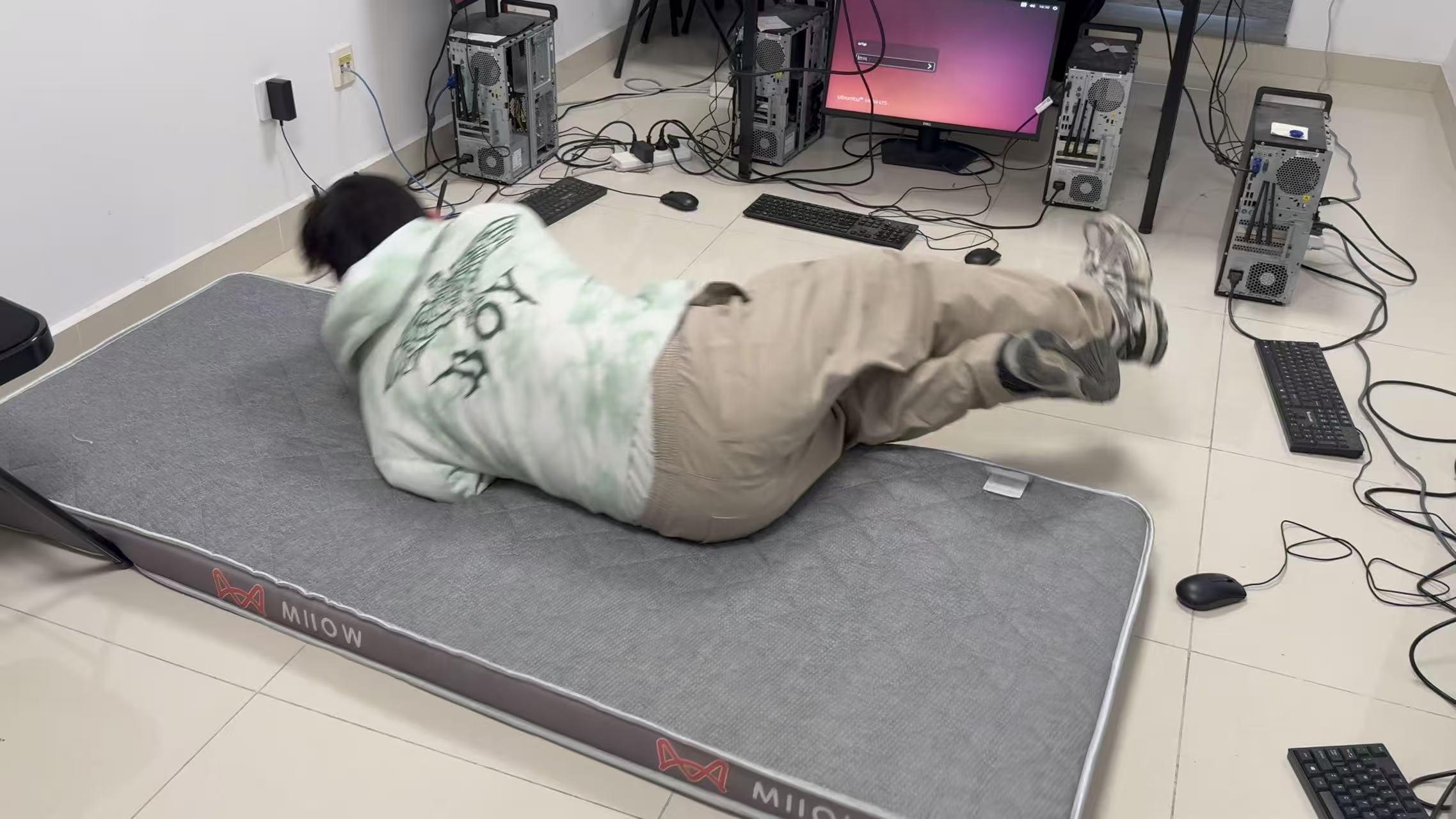}
			\label{fig_third_case}}
		\hfil
		\subfloat[]{\includegraphics[width=0.4\linewidth]{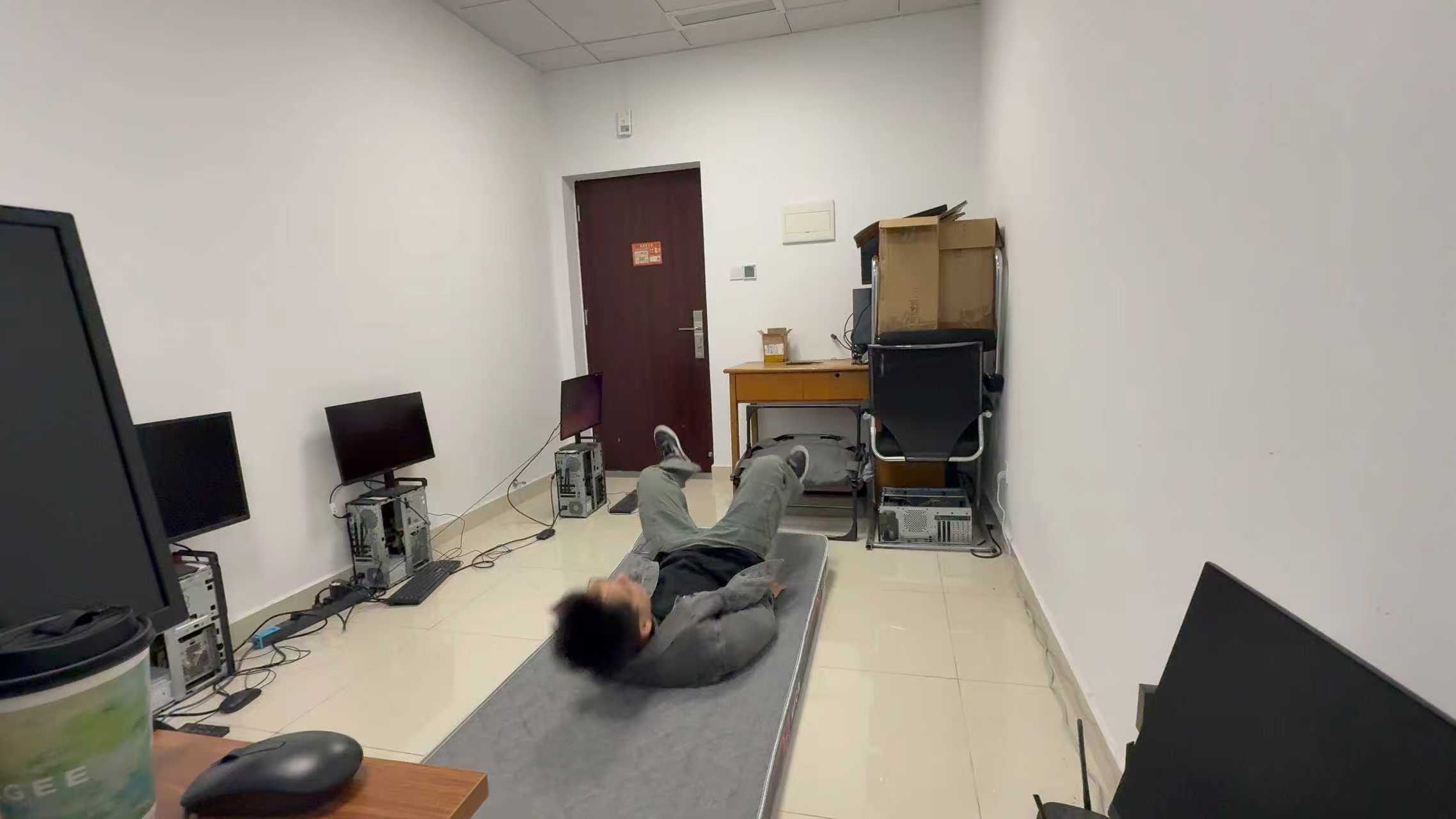}
			\label{fig_four_case}}		
		\caption{Falls including forward, lateral, and backward falls}
		\label{fig_sim}
	\end{figure}

	We deploy the system in a completely novel environment (a typical office room filled with a few clutter) never exposed to the model during training which is shown in Fig \ref{room}.  We test falls from different directions with
	different persons in this environment which is shown in Fig \ref{fig_sim}. During continuous live tests involving various daily activities and simulated falls which is shown in Table \ref{tab:event_accuracy}, the system maintain an empirical accuracy exceeding 90\%, with near-zero false alarms triggered by normal activities.
	
	\begin{table}[htbp]
		\caption{Number of events tested}
		\label{tab:event_accuracy}
		\centering
		\begin{tabular*}{\columnwidth}{@{\extracolsep{\fill}}lcc@{}}
			\toprule
			\textbf{Events} & \textbf{Number} & \textbf{Accuracy} \\
			\midrule
			Fall forward  & 15  & 100\% \\
			Fall lateral  & 20  & 95\% \\
			Fall backward & 15  & 93.3\% \\
			Sit down and stand up      & 15  & 86.7\%  \\
			Pick up an object          & 10  & 90\%  \\
			Walk          & 15  & 93.3\%  \\
			Wave the arm  & 10  & 100\%  \\
			\midrule
			Total         & 100 & 94\% \\
			\bottomrule
		\end{tabular*}
	\end{table}
	\subsubsection{Latency and IoT Feasibility}
	For time-critical applications like fall detection, latency is paramount. On an edge device equipped with an entry-level GPU, the end-to-end processing pipeline incurred an average inference latency of merely 12.4 ms per window. The network comprises approximately 10.5 million parameters. During inference on an NVIDIA GPU, the model achieves a processing speed of 648 frames/samples per second (FPS), with an average inference time of just 1.5 ms per CSI spectrogram segment. This confirms that despite the integration of dual-attention mechanisms and Transformer modules, the architecture maintains high efficiency, fully satisfying the stringent low-latency requirements for real-time indoor safety monitoring.
	
	\section{Conclusion}
	In this paper, we presented a robust, domain-generalizable WiFi fall detection framework to overcome the critical bottlenecks of environmental domain shifts and NLoS signal attenuation. We proposed an Attention-Enhanced CNN-Transformer architecture featuring a novel physics-driven DVG, coupled with a Physics-Aware Data Augmentation strategy. Specifically, the DVG acts as a pre-processing attention mechanism to dynamically decouple the static environmental background from human motion signatures. By leveraging Instance Normalization and simulating physical signal variances (e.g., distance-induced scaling and wall-induced low-pass filtering), the network is strictly forced to learn invariant morphological fall features. Furthermore, integrating the CBAM enables the model to dynamically suppress residual background noise and localize critical spatiotemporal features. Building upon this spatial-channel refinement, a Transformer Encoder is utilized to capture the long-range temporal dependencies and causal relationships inherent in the multi-stage fall process, ensuring accurate discrimination between genuine falls and complex confounding activities. 
	
	Extensive cross-domain evaluations demonstrated our framework's superiority, achieving up to 97.6\% accuracy in challenging NLoS scenarios and a resilient 98.8\% accuracy in completely unseen environments without target-domain fine-tuning. Crucially, real-world live inference field tests on an edge computing system confirmed its low-latency performance and practical robustness against physical environmental changes. Ultimately, this work provides a reliable, privacy-preserving solution for continuous, whole-home elderly monitoring.

	\bibliographystyle{IEEEtran}
	\bibliography{refs}
	
	
	
%
%
%
	
\end{document}